\documentclass[preprint,aps,pre,epsf, superscriptaddress]{revtex4}
\usepackage{amsmath}
\usepackage{amssymb}
\usepackage{epsfig}
\usepackage{graphicx}
\usepackage{multirow}
\usepackage[T1]{fontenc}
\usepackage[utf8]{inputenc}
\usepackage{color}
\usepackage[normalem]{ulem}
\usepackage{tikz}
\usetikzlibrary{positioning}
\usepackage{tabularx}

\graphicspath {{./figures/}}
\makeatletter
\def\input@path{{./figures/}}
\makeatother

\begin{document}

\title{Disorder effects on the metastability of classical Heisenberg ferromagnets}

\author{Moumita Naskar}
\affiliation{Department of Physics, Presidency University,
86/1 College Street, Kolkata-700073, India}

\author{Muktish Acharyya}
\email{muktish.physics@presiuniv.ac.in}
\affiliation{Department of Physics, Presidency University,
86/1 College Street, Kolkata-700073, India}

\author{Erol Vatansever}
\affiliation{Department of Physics, Dokuz Eyl\"{u}l University, TR-35160, Izmir, Turkey}
\affiliation{Centre for Fluid and Complex Systems, Coventry
	University, Coventry, CV1 5FB, United Kingdom}
 
\author{Nikolaos G. Fytas}
\email{nikolaos.fytas@coventry.ac.uk}
\affiliation{Centre for Fluid and Complex Systems, Coventry
	University, Coventry, CV1 5FB, United Kingdom}
\date{\today}

\begin{abstract}
In the present work, we investigate the effects of disorder on the reversal time ($\tau$) of classical anisotropic Heisenberg ferromagnets in three dimensions by means of Monte Carlo simulations. Starting from the pure system, our analysis suggests that $\tau$ increases with increasing anisotropy strength. On the other hand, for the case of randomly distributed anisotropy, generated from various statistical distributions, a set of results is obtained: (i) For both bimodal and uniform distributions the variation of $\tau$ with the strength of anisotropy strongly depends on temperature. (ii) At lower temperatures, the decrement in $\tau$ with increasing width of the distribution is more prominent. (iii) For the case of normally distributed anisotropy, the variation of $\tau$ with the width of the distribution is non-monotonic, featuring a minimum value that decays exponentially with the temperature. Finally, we elaborate on the joint effect of longitudinal ($h_z$) and transverse ($h_x$) fields on $\tau$, which appear to obey a scaling behavior of the form $\tau h_z^{n} \sim f(h_x)$.
\end{abstract}

\maketitle

\section{Introduction}
\label{sec:introduction}

The metastable behavior of a ferromagnet is an interesting field of modern research with important technological applications~\cite{gunton83,rikvold94}. It is well-known today that the so-called switching time (or reversal time $\tau$), a critical parameter of this process, plays a key role in the speed of recording in magnetic storage devices~\cite{chong12}. The whole problem of metastability dates back to 1935, where the classical theory of nucleation was developed by Becker and D\"oring~\cite{becker35}, predicting the growth of supercritical droplets. These predictions of the different regimes of such growth depending on the magnitude of the applied magnetic field were successfully verified by extensive Monte Carlo simulations in Ising~\cite{stauffer99} and Blume-Capel ferromagnets~\cite{naskar21}, as well as in generalized spin-$s$ anisotropic models~\cite{naskar21b}. 

Over the years, most numerical approaches of reversal phenomena dealing with the growth of supercritical clusters, the decay of metastable volume fraction, and their dependencies on the applied field and temperature, focused on pure ferromagnetic systems of discrete symmetric Ising and Blume-Capel models. A straightforward extension would be then to consider the case of continuous symmetric spin models, such as the classical Heisenberg model. Indeed, a few papers have already shed some light in this context: The magnetization switching in the classical anisotropic Heisenberg ferromagnet was studied by Monte Carlo simulations and the dynamics of coherent spin rotation was detected in the limit of low anisotropy~\cite{hinzke98}. Later on, the reversal properties were studied in classical anisotropic ferromagnetic and antiferromagnetic bilayer Heisenberg models and it was observed that the magnetization behavior is different at each branch of the hysteresis loop as well as the exchange-bias behavior~\cite{santos14}. More recently, the problem was investigated in the antiferromagnetic anisotropic Heisenberg chain~\cite{kolesnikov17} and in a Van-der-Waals magnet where a strain-sensitive magnetization reversal was reported~\cite{wang20}. This latter work indicated that lattice deformation plays a major role in the reversal process, as it may lead to random variations of the crystal field (anisotropy) acting on the spins of the system.

In all aforementioned studies regarding the anisotropic classical Heisenberg ferromagnet, uniform anisotropy was used for simplicity. At this point several intriguing questions may be raised: (i) What will be the effect of a random anisotropy on the reversal phenomena in the classical Heisenberg ferromagnet? (ii) How does the nature of the statistical distribution of the anisotropy affect the reversal of magnetization and other properties of the system? To the best of our knowledge, all these open fundamental aspects have not yet been addressed so our understanding of metastable phenomena in a random environment is rather limited -- see Refs.~\cite{naskar20,mandal21} for some particular exceptions. In the present work, we make one step forward in this direction by studying via Monte Carlo simulations the three-dimensional anisotropic classical Heisenberg ferromagnet with the disorder, generated by a set of random anisotropy distributions.

The rest of the paper is organized as follows: Section~\ref{sec:model} provides a description of the model together with an outline of the simulation scheme. The numerical results and scaling analysis are reported in Sec.~\ref{sec:results}. This paper concludes with a summary and outlook in Sec.~\ref{sec:conclusions}.

\section{Model and Numerics}
\label{sec:model}

The classical anisotropic (uniaxial and single-site) Heisenberg model with nearest-neighbor ferromagnetic interactions in the presence of an external magnetic field is described by the following Hamiltonian
\begin{equation}
\label{eq:Ham}
\mathcal {H} = -J \sum_{\langle ij\rangle} {\bf S_i} \cdot {\bf S_j} - \sum_i D_i \; (S_i^z)^2 - {\bf h} \sum_i {\bf  S_i},
\end{equation}
where ${\bf S_i}\;(S_i^x, S_i^y, S_i^z)$ represents a classical spin vector with $|{\bf S}|=1$ (or $(S_i^x)^2+(S_i^y)^2+(S_i^z)^2 = 1$) which is allowed to take any (unrestricted) angular orientation in the vector space. The first term in the Hamiltonian corresponds to nearest-neighbor ($\langle ij\rangle$) ferromagnetic ($J > 0$) spin-spin interaction. The parameter $D_i$ appearing in the second term denotes the strength of uniaxial anisotropy favoring the $z$-axis alignment of the spin vector. Note that the limit $D \rightarrow \infty$ corresponds to the Ising ferromagnet, whereas for $D = 0$ Eq.~(\ref{eq:Ham}) reduces to the isotropic Heisenberg Hamiltonian. Finally, the last term in the Hamiltonian~(\ref{eq:Ham}) stands for the interaction of individual spins with the externally applied magnetic field vector ${\bf h}\;(h_x, h_y, h_z)$. 
The reversal of magnetization is studied mainly in the presence of a longitudinal field $h_z$ unless otherwise stated. However, in the last part of our study, we also present the influence of an additional transverse field ($h_x$) on the reversal mechanism, along with the longitudinal field.

We used Monte Carlo simulations of Metropolis type to study the model of Eq.~(\ref{eq:Ham}) on simple cubic lattices with periodic boundary conditions and linear dimension $L$, where typically $L = 50$ (more information on the numerical details is given in Appendix~\ref{numerics}). In all our numerical experiments, $L^3$ such spin updates define one Monte Carlo step per site, which also sets the time unit of our simulations. 
We also fix $J = k_{\rm B} = 1$ to properly set the temperature scale. 

Finally some comments about errors and fitting analysis: Unless otherwise stated, we always perform the necessary statistical averaging to increase the accuracy of our data (more details are given explicitly in the plots of Sec.~\ref{sec:results}) and compute standard errors~\cite{barkema99}. For the fits, we implement the $\chi^2$ test of goodness of fit~\cite{press92}. Specifically, the $Q$ value of our $\chi^2$ test is the probability of finding a $\chi^2$ value that is even larger than the one actually found from our data. We consider a fit as being fair only if $10\% < Q < 90\%$. 

\section{Results and discussion}
\label{sec:results}

\begin{figure}[htbp]
	\includegraphics[width=75mm]{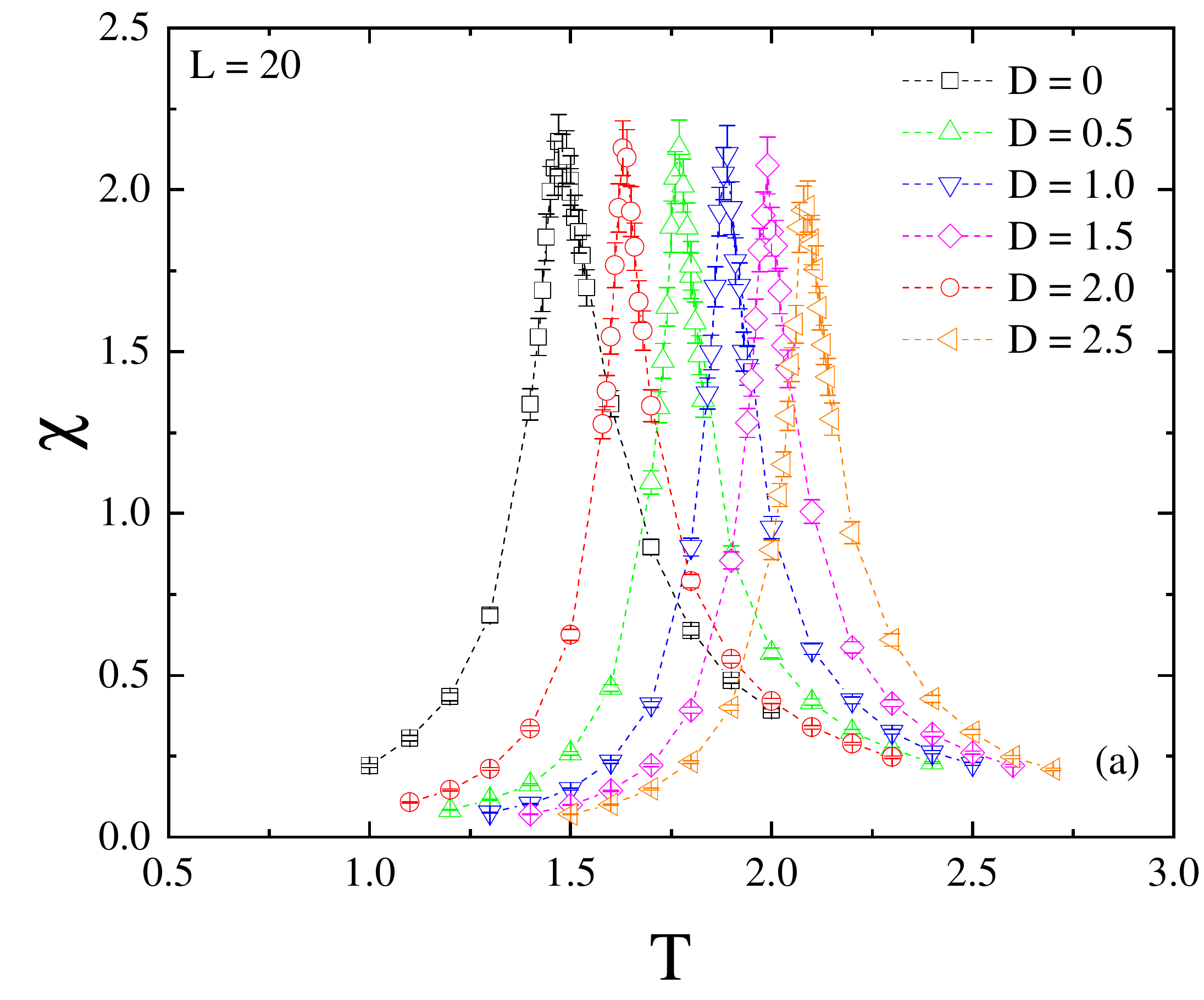}
	\includegraphics[width=80mm]{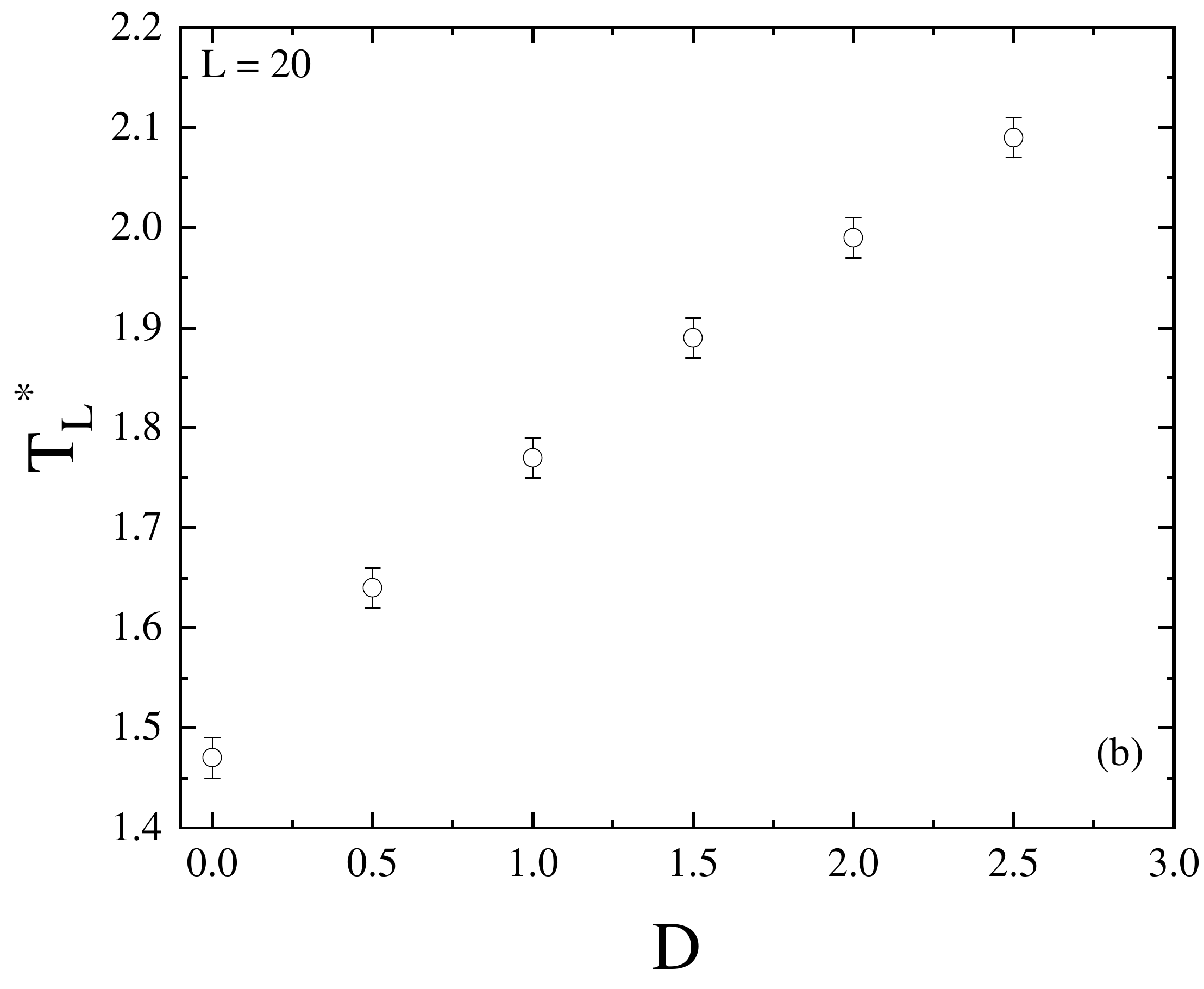}
	\caption{\label{tc}
		(a) Magnetic susceptibility ($\chi$) vs. temperature ($T$) for different strengths of the uniform anisotropy $D$. (b) Corresponding pseudo-critical temperatures $T_{L}^{\ast}$ vs. $D$, as obtained from panel (a). The temperature is varied in steps of $10^{-1}$ when obtaining the magnetic susceptibility curves, so the maximum error in the determination of each $T_L^*$ is considered to be $2\times 10^{-2}$.} Results are shown for a system with linear size $L=20$ averaged over $10$ samples.
\end{figure}

The first port of call in our study is to secure a rough estimate for the critical temperature of the Heisenberg model so that we assure that the system lies well below its critical temperature. 
Although in previous works~\cite{naskar211} we observed no such detectable discontinuity in the behavior of the metastable lifetime across the phase boundary of the ferromagnetic-paramagnetic transition, still, in the present study, we follow the standard way of studying the reversal time or the metastable lifetime below the critical temperature. As the precise determination of the critical temperature is definitely not necessary here, we follow the simplest practice for locating an approximate estimation of the critical temperature. The method refers to the detection of the pseudo-critical temperature, $T_{L}^{\ast}$, at which the magnetic susceptibility $\chi$ obtained via
\begin{equation}
\label{eq:chi}
\chi = N \beta \big[\langle m^2 \rangle - \langle m \rangle ^2\big]
\end{equation}
attains a peak for a relatively large system size $L$. In the above Eq.~(\ref{eq:chi}), $N = L^3$, $\beta=1/T$, and $m=\sqrt{m_x^2+m_y^2+m_z^2}$ denotes the magnetization of the system, where $m_x=(\sum_i S_i^x) / N$, $m_y=(\sum_i S_i^y) / N$, and $m_z=(\sum_i S_i^z) / N$. Figure~\ref{tc}(a) presents the $T$-variation of $\chi$ for a system with linear size $L=20$ and for various values of $D$ and Fig.~\ref{tc}(b) clearly illustrates that the pseudo-critical temperatures obtained increase with increasing uniform anisotropy $D$. Note that the temperature is varied in steps of $\delta T = 10^{-2}$ so that the maximum error in the $T_{L}^{\ast}$-determination is $\delta T_{L}^{\ast}\sim 2.10^{-2}$. 

\begin{figure}[htbp]
	\includegraphics[width=80mm]{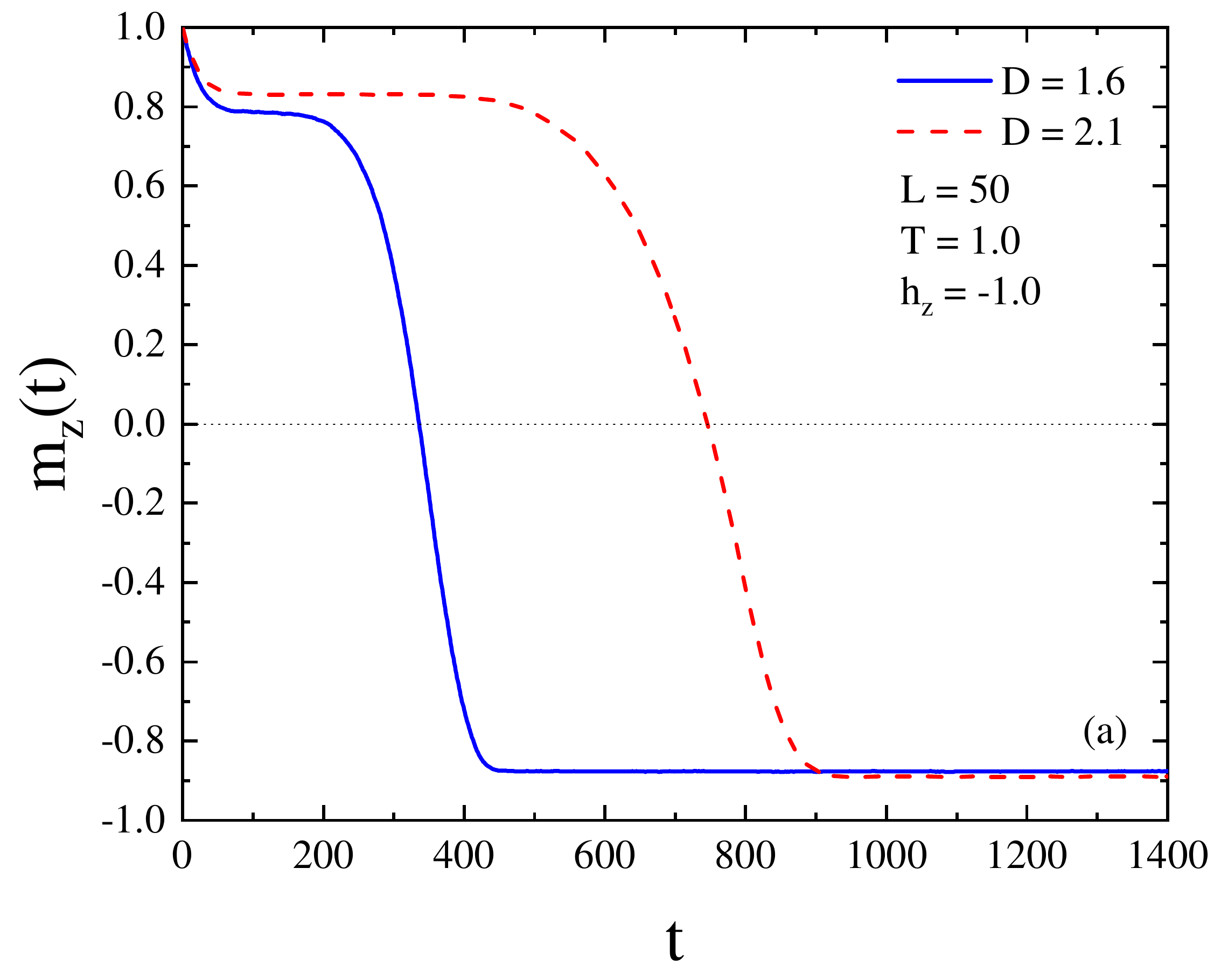}\\
	\includegraphics[width=80mm]{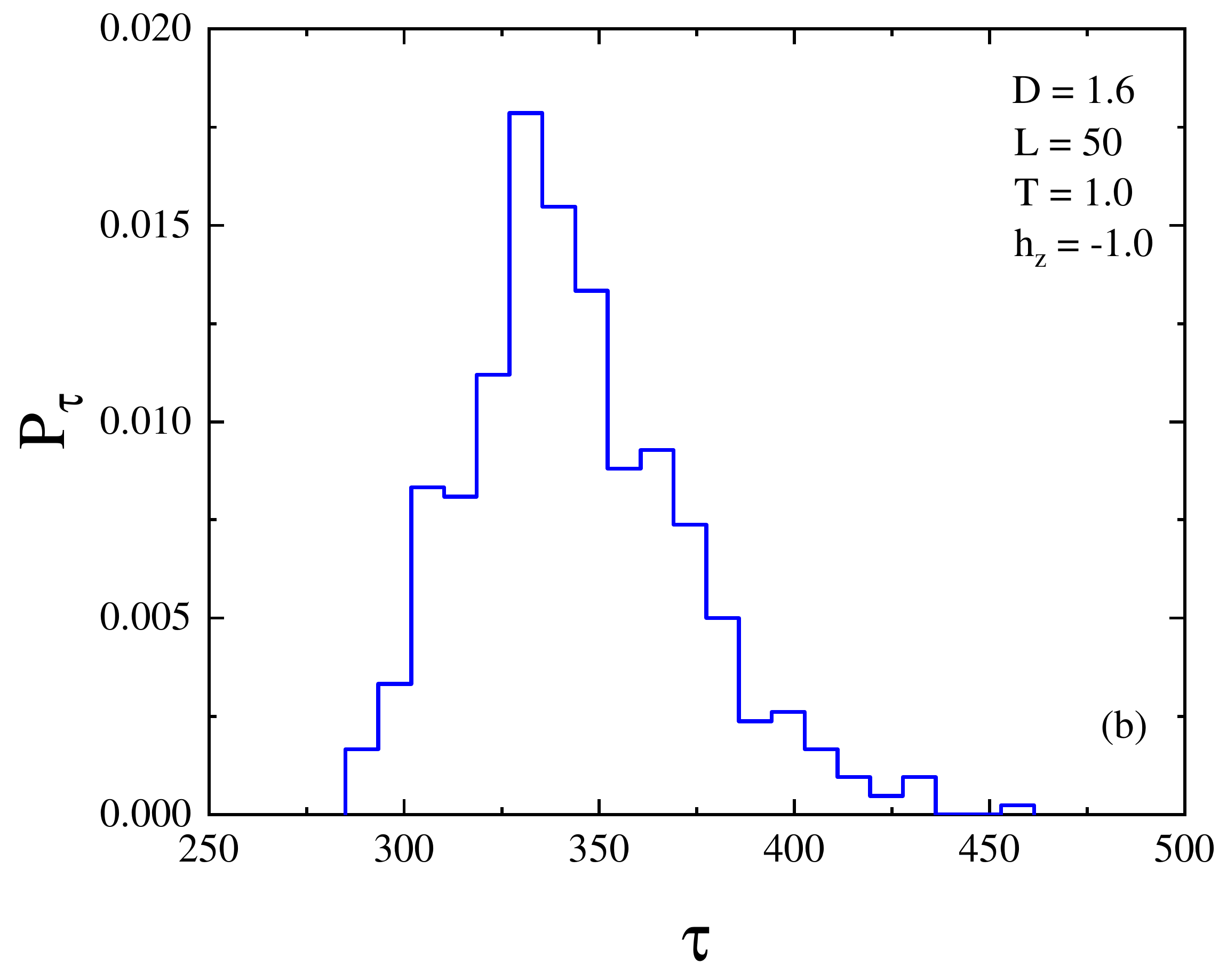}\\
        \includegraphics[width=80mm]{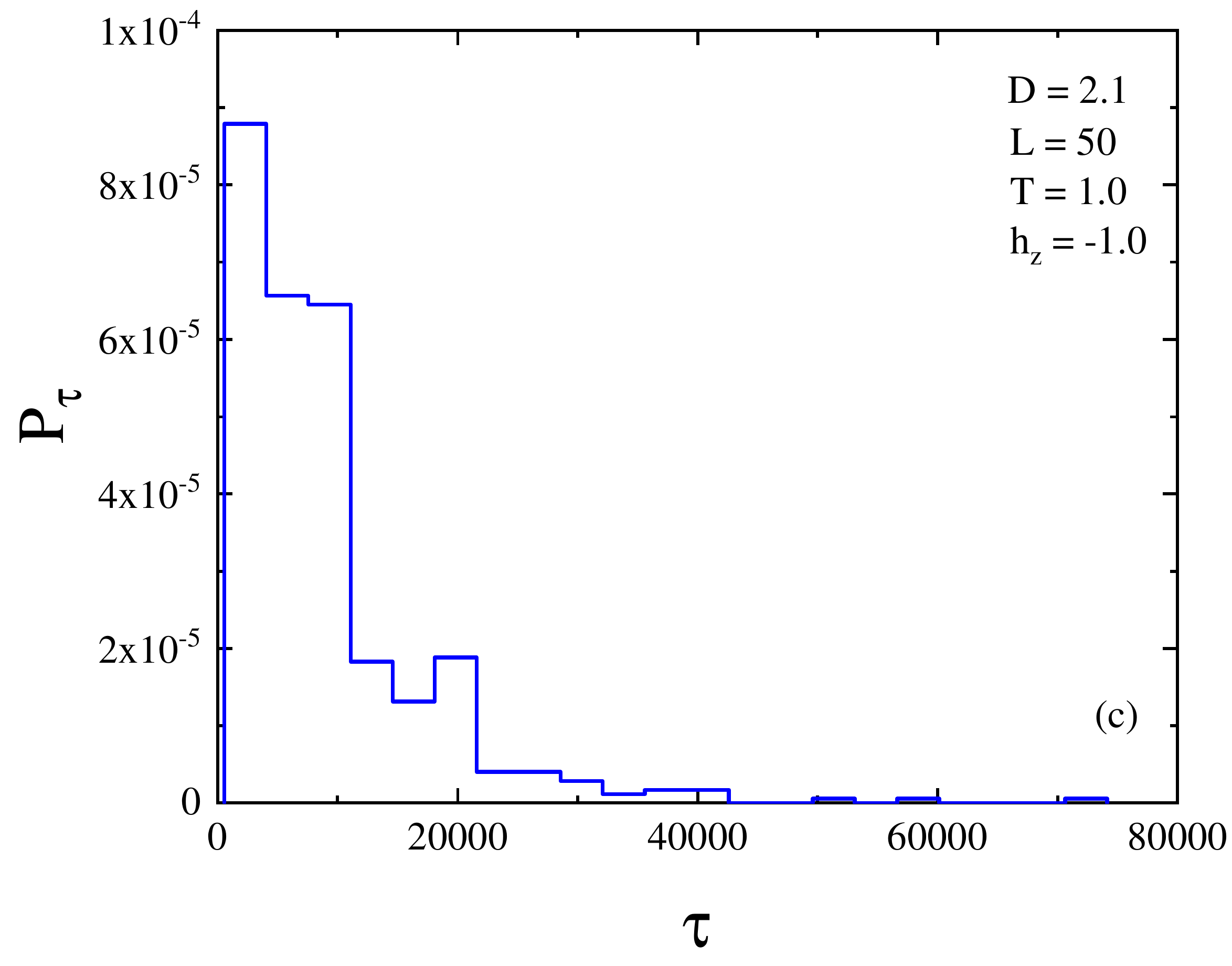}
	\caption{\label{tau_D-dist}
		(a) Magnetization ($m_{z}$) vs. time ($t$) of a single sample for two different strengths of the uniform anisotropy, $D = 1.6$ and $D = 2.1$. Normalized probability distribution of reversal times $P_{\tau}$ over $500$ samples for $D = 1.6$ (b) and $D = 2.1$ (c). The simulation parameters in all panels are $L=50$, $h_z=-1.0$, and $T = 1.0$.}
\end{figure}

The metastable lifetime (or reversal time) $\tau$, the crucial parameter of our analysis, is defined quantitatively as the time at which the magnetization along the $z$-direction first changes sign ($m_z \simeq 0$). A typical decay of a metastable state for a single sample is presented in Fig.~\ref{tau_D-dist}(a) for two different strengths of uniform anisotropy, $D = 1.6$ and $2.1$. Correspondingly, Figs.~\ref{tau_D-dist}(b) and (c) present the statistical distributions of $\tau$, $P_{\tau}$, obtained over $500$ different samples. These panels suggest that the standard deviation of the distributions increases with increasing anisotropy. Hereafter, the parameter of interest will be the mean metastable lifetime obtained by taking a simple arithmetic average over the ensemble.

The effects of uniform anisotropy but also uniform random anisotropy on the variation of the reversal time are sketched in Fig.~\ref{tau-D}. Let us clarify here the three cases under study:

(i) The first case refers to uniform anisotropy, where $D_i = D \; \forall i$, so that each lattice site experiences the same strength of anisotropy $D$.

(ii) The second case refers to the uniform random distribution of the anisotropy, as obtained from the probability distribution $\mathcal{P}_{\rm u}(D_i)= 1/W_{\rm DU}$ with mean value $\mu=0$. For such distribution, the strength of the anisotropy is randomly and uniformly distributed between $\{-W_{\rm DU}/2, W_{\rm DU}/2\}$ over the lattice sites.

(iii) In the third case, we consider the combination of the above two cases, \emph{i.e.}, the resultant of uniform anisotropy and uniformly distributed random anisotropy with $\mathcal{P}'_{\rm u}(D_i)= 1/W_{\rm DU}$ and a shifted mean value of $\mu = D$ in this case. Here the anisotropy is randomly and uniformly distributed between $\{W_{\rm DU}/2, 3W_{\rm DU}/2\}$ over the lattice sites. 

In Fig.~\ref{tau-D}(a) the first and third cases outlined above are represented by open and filled symbols while the second case is denoted by red stars, respectively. The curves corresponding to the first and third cases are obtained for two different temperature values, $T = 0.8$ and $1.0$ and the second case is investigated only for $T = 1.0$.  In the first case, we observe that the reversal time increases with the increasing strength of the uniform anisotropy $D$. This is due to the fact that, as $D$ increases, the alignment of spins along the $z$-axis becomes energetically more favorable and thus the energy barrier of the metastable state increases leading to a longer metastable lifetime. Remarkably, we find here the presence of a crossover analogous to a similar type of behavior reported in our previous study of the reversal time with respect to the inverse of the applied field in the spin-$1/2$ Ising system~\cite{naskar21b}. For the second case and interestingly enough, when a uniform random distribution of anisotropy with a mean value $\mu = 0$ is considered (red stars), the system appears to mimic the behavior of an isotropic system with $D = 0$. On the contrary, for the case of a uniform random anisotropy with mean value $\mu = D$ (green triangles and blue circles), the reversal time varies similarly to the first case of a uniform anisotropy. At this point it worth noting that we have also performed some additional test simulations to probe any possible finite-size effects on the variation of the reversal time in the presence of uniform crystal field coupling. We refer the reader to the discussion in Appendix~\ref{finite-size-effects}). 

In Fig.~\ref{tau-D}(b) now we focus explicitly on the case (ii) and study the system for a set of different temperatures. From the numerical data, it is clear that there is a temperature dependence on the behavior of the metastable lifetime. 
At high temperatures the lifetime is unaffected, resembling the behavior of the isotropic system. However, at low temperatures, the reversal time decreases with the increasing strength of the distribution. 
Thus, as also noted in the previous paragraph, the deviation in Fig.~\ref{tau-D}(a) is expected to be enhanced at $T=0.4$ when compared to that at $T=1.0$.

\begin{figure}[htbp]
	\includegraphics[width=79mm]{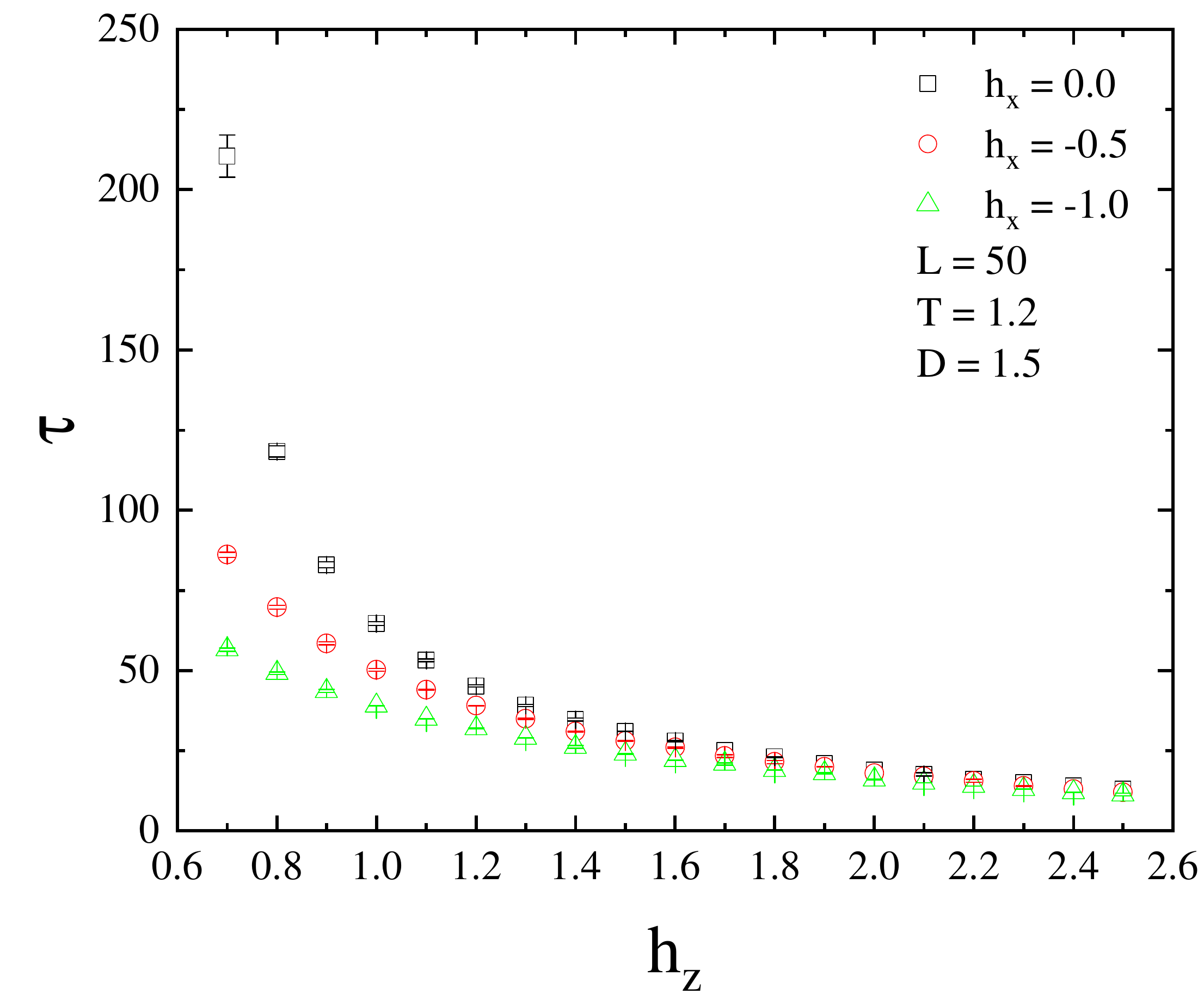}
	\caption{\label{long-h}
		Effect of the transverse field $h_x$ on the variation of reversal time with longitudinal field $h_z$ for a system with linear size $L = 50$ at $D=1.5$ and $T=1.2$. Results averaged over $20$ samples.}
\end{figure}

\begin{figure}[htbp]
	\includegraphics[width=79mm]{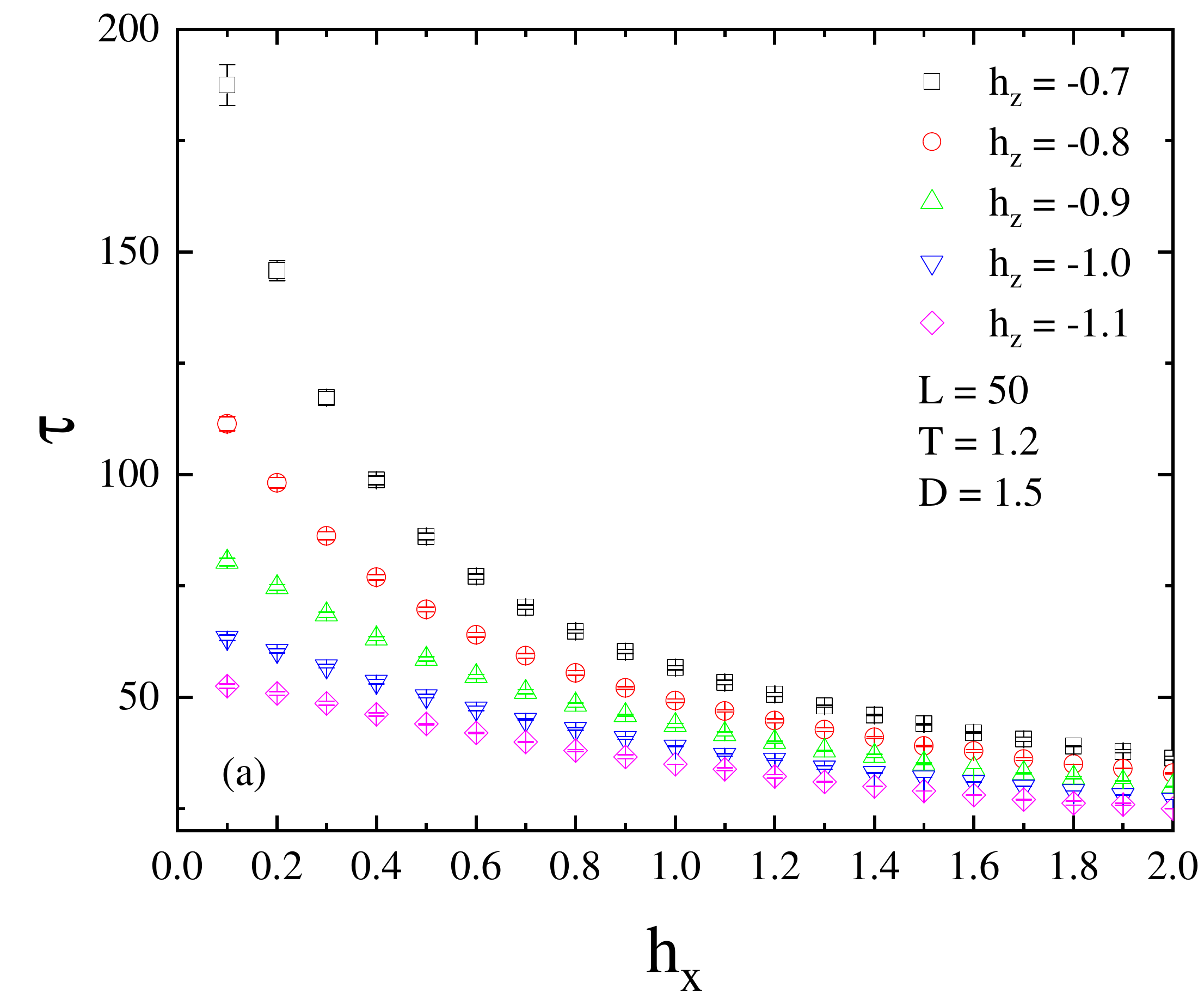}
        \includegraphics[width=80mm]{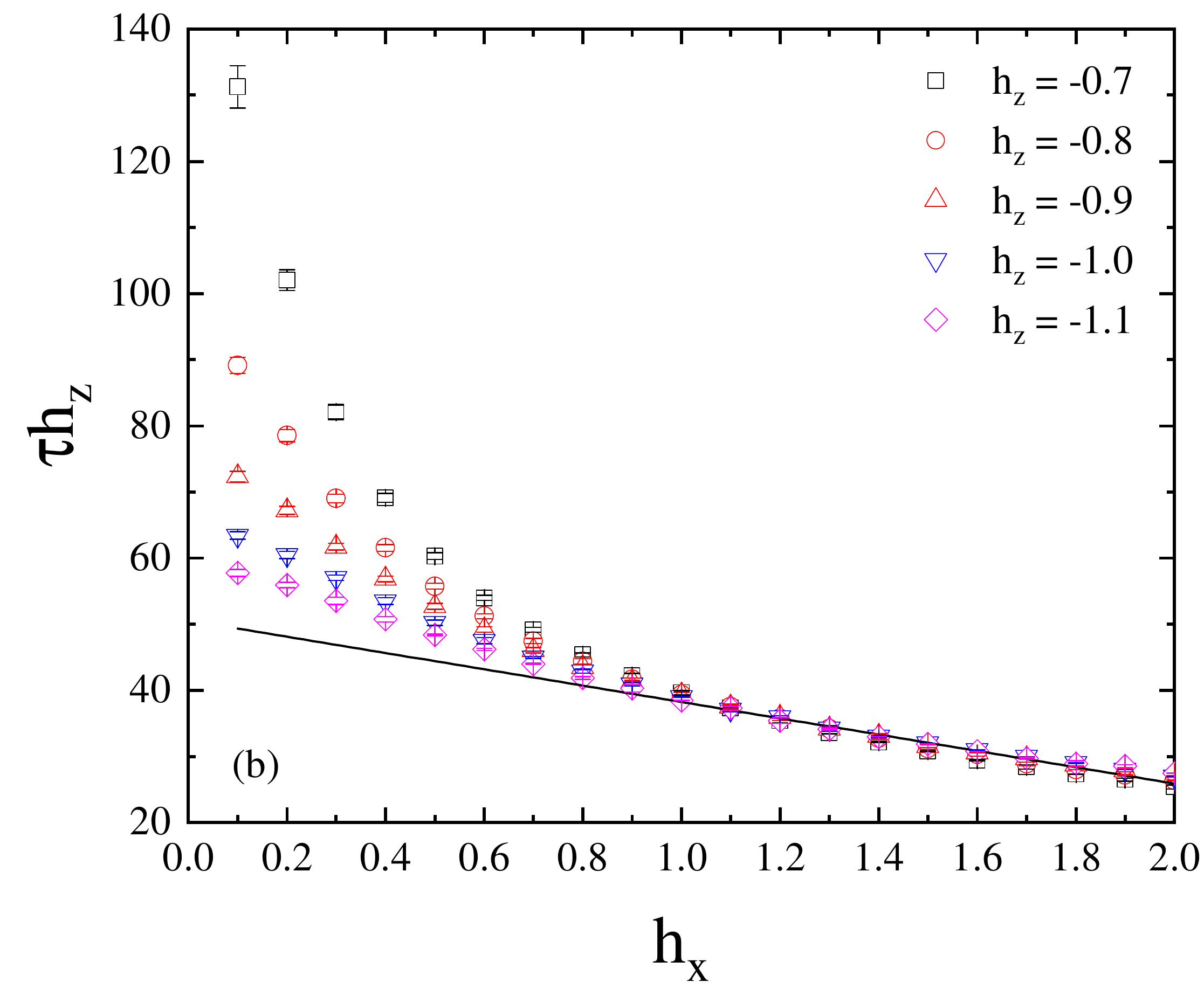}
	\caption{\label{datacollapse}
		Effect of the longitudinal field $h_z$ on the variation of reversal time with transverse field $h_x$ (a) and the corresponding data collapse (b) using the scaling law discussed in the main text. The solid line shows a linear fit in the strong $h_x$ regime. The system's parameters are identical to those of Fig.~\ref{long-h}.}
\end{figure}

At this point and before investigating the effects of other disorder distributions on the reversal time of the magnetization, we report here the behavior of $\tau$ in the presence of both longitudinal and transverse fields for the case of uniform anisotropy. For a fixed value of the transverse field we find that the reversal time decreases with increasing strength of the longitudinal field; see Fig.~\ref{long-h}. Furthermore, with the application of a transverse field alongside the longitudinal field, reversal also becomes faster, as depicted in Fig.~\ref{datacollapse}(a). For this particular case, a scaling law of the form $\tau = h_z^{-n} f(h_x)$ with  $n = 1$ appears to adequately describe the collapse of the data, see Fig.~\ref{datacollapse}(b), at least within the strong $h_x$ regime.

\begin{figure}[htbp]
	\includegraphics[width=80mm]{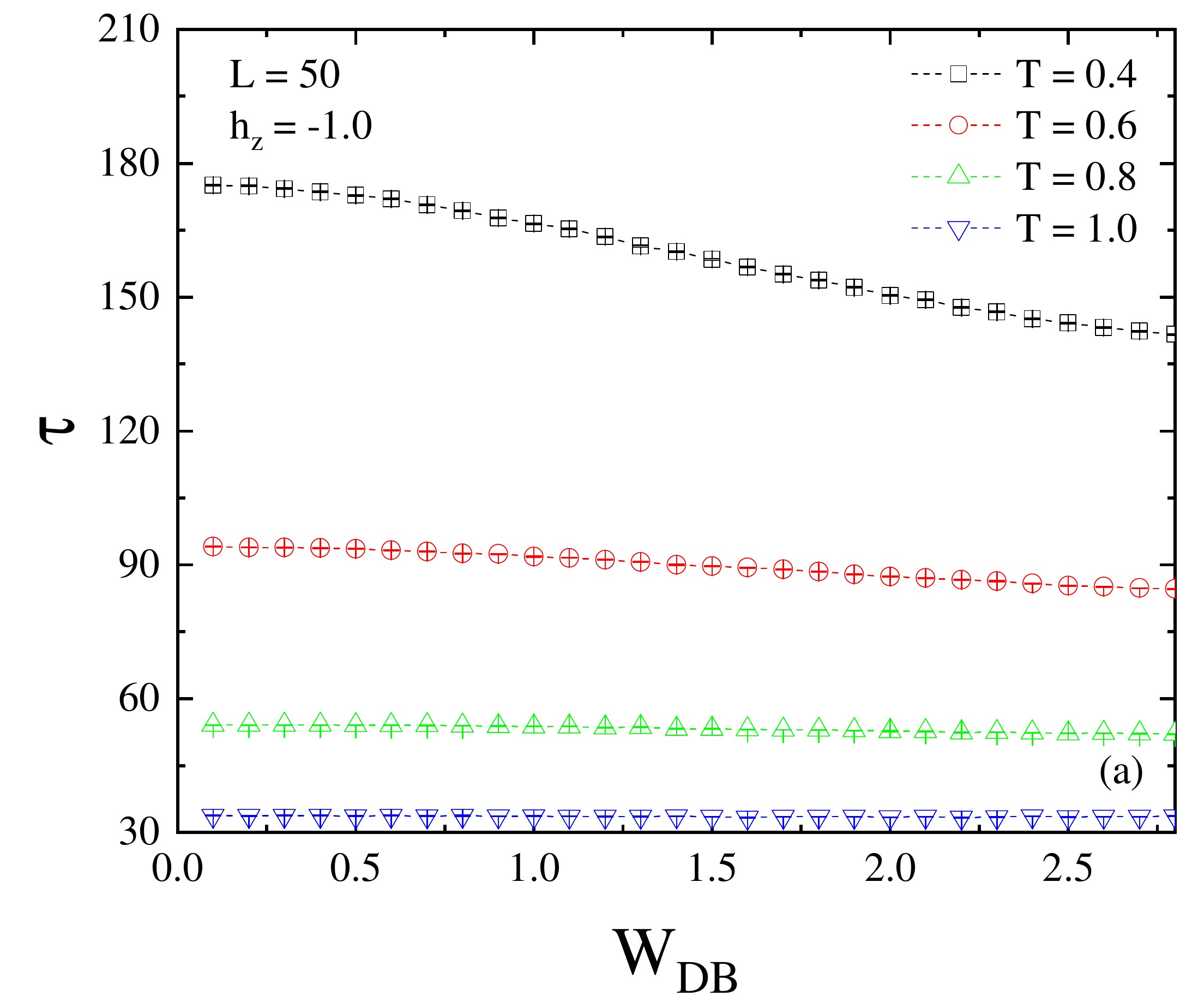}
	\includegraphics[width=82mm]{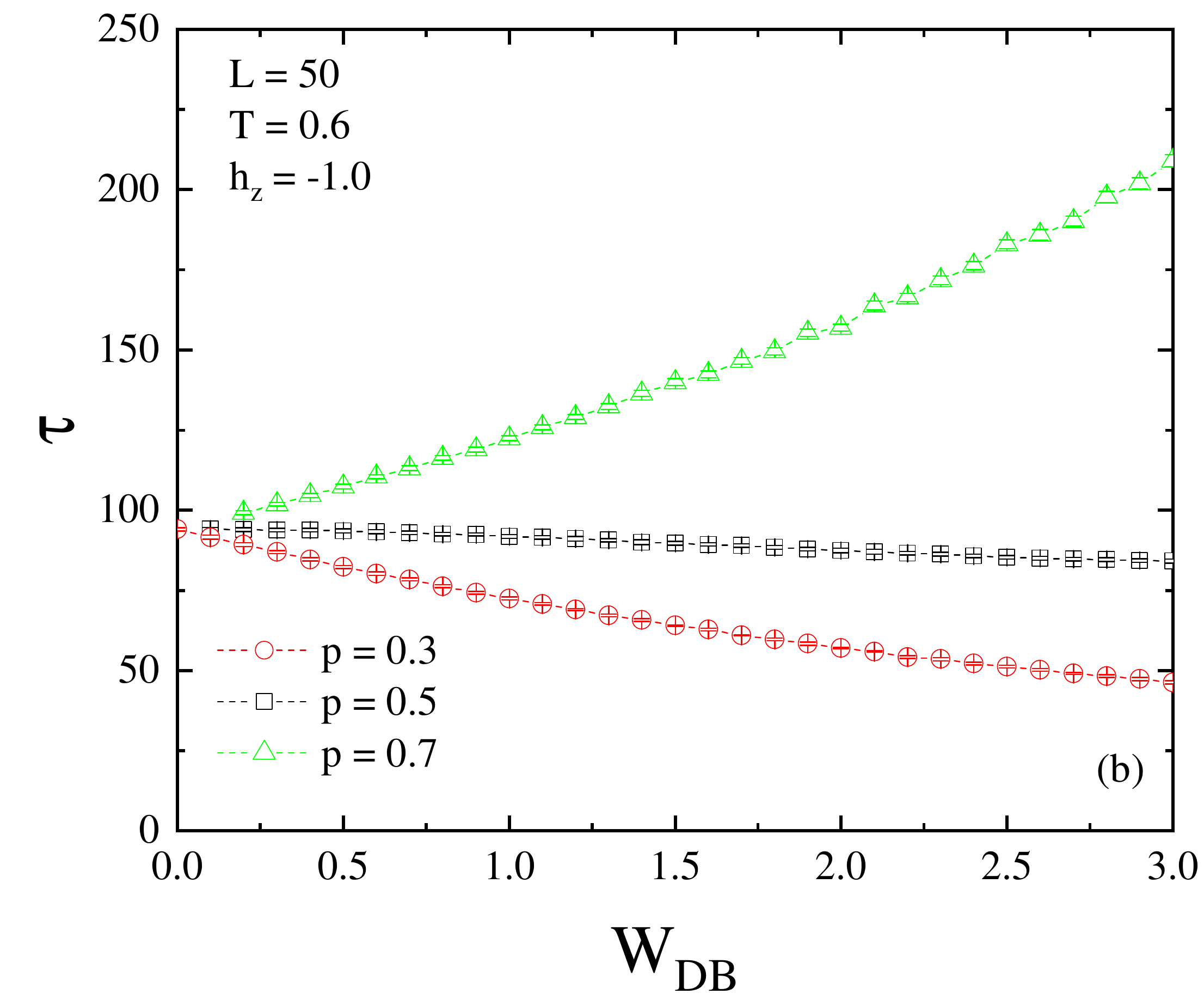}
	\caption{\label{tau_D-bi}
		(a) Reversal time vs. the strength of the bimodal distribution of anisotropy~(\ref{eq:bimodal}) for the case $p = 0.5$ and various temperatures in the range $T=0.4-1.0$. (b) Similar to panel (a) but for three versions of the bimodal distribution with $p = 0.3$ (red circles), $p = 0.5$ (black squares), and $p = 0.7$ (green triangles) at $T = 0.6$. All results correspond to a system with linear size $L = 50$, $h_{z} = -1$, and were averaged over $80$ realizations.}
\end{figure}

In analogy to Fig.~\ref{tau-D}, we present in Fig.~\ref{tau_D-bi} the variation of the metastable lifetime of the system, now in the presence of a bimodal distribution of anisotropy of strength $W_{\rm DB}$, obtained from the well-known distribution 
\begin{equation}
\label{eq:bimodal}
\mathcal{P}_{\rm B}(D_i)= p\delta (D_i - \frac{W_{\rm DB}}{2}) + (1-p)\delta (D_i + \frac{W_{\rm DB}}{2})
\end{equation}
at various temperatures. In particular, in Fig.~\ref{tau_D-bi}(a) we consider the most symmetric case with $p = 0.5$. Here also, the reversal time appears to decrease with increasing strength of the bimodal distribution at lower temperatures. A simple modulation of the bimodal distribution, for instance, $p = 0.3$, reduces the reversal time even further, as the mean of the distribution is shifted to negative values; red circles in Fig.~\ref{tau_D-bi}(b). An opposite scenario is observed for $p = 0.7$ as can be seen from the relevant data in the same panel.

\begin{figure}[htbp]
	\includegraphics[width=82mm]{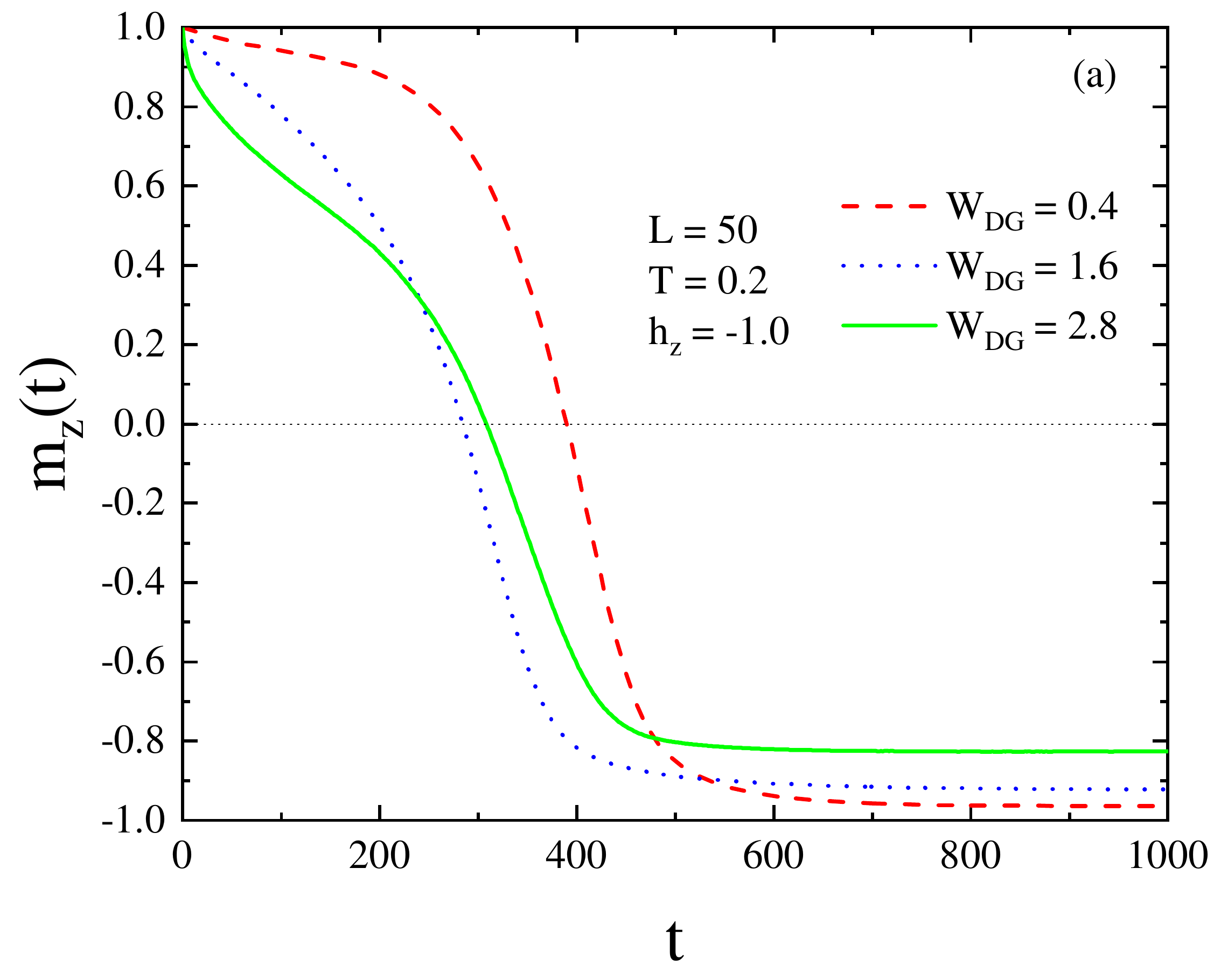}
	\includegraphics[width=79mm]{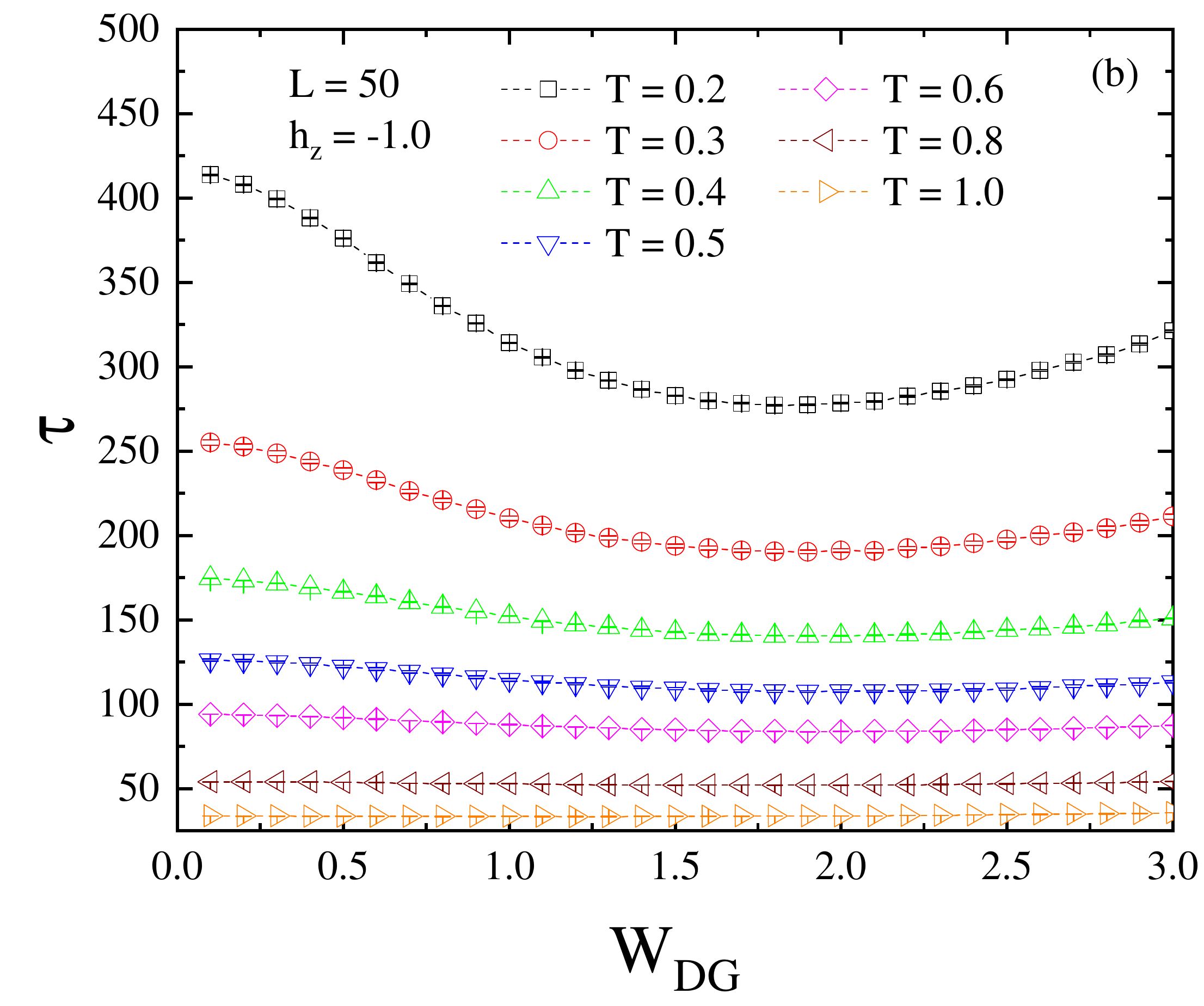}
	\caption{\label{tau_D-Gauss}
		(a) Magnetization ($m_{z}$) vs. time ($t$) of a typical single sample for three different strengths of the Gaussian distribution of anisotropy~(\ref{eq:gaussian}). (b) Reversal time vs. the strength of the Gaussian distribution at different temperatures within the range $T = 0.2 - 1.0$. All results correspond to a system with linear size $L = 50$, $h_{z} = -1$, and were averaged over $50$ realizations.}
\end{figure}

\begin{figure}[htbp]
	\includegraphics[width=79mm]{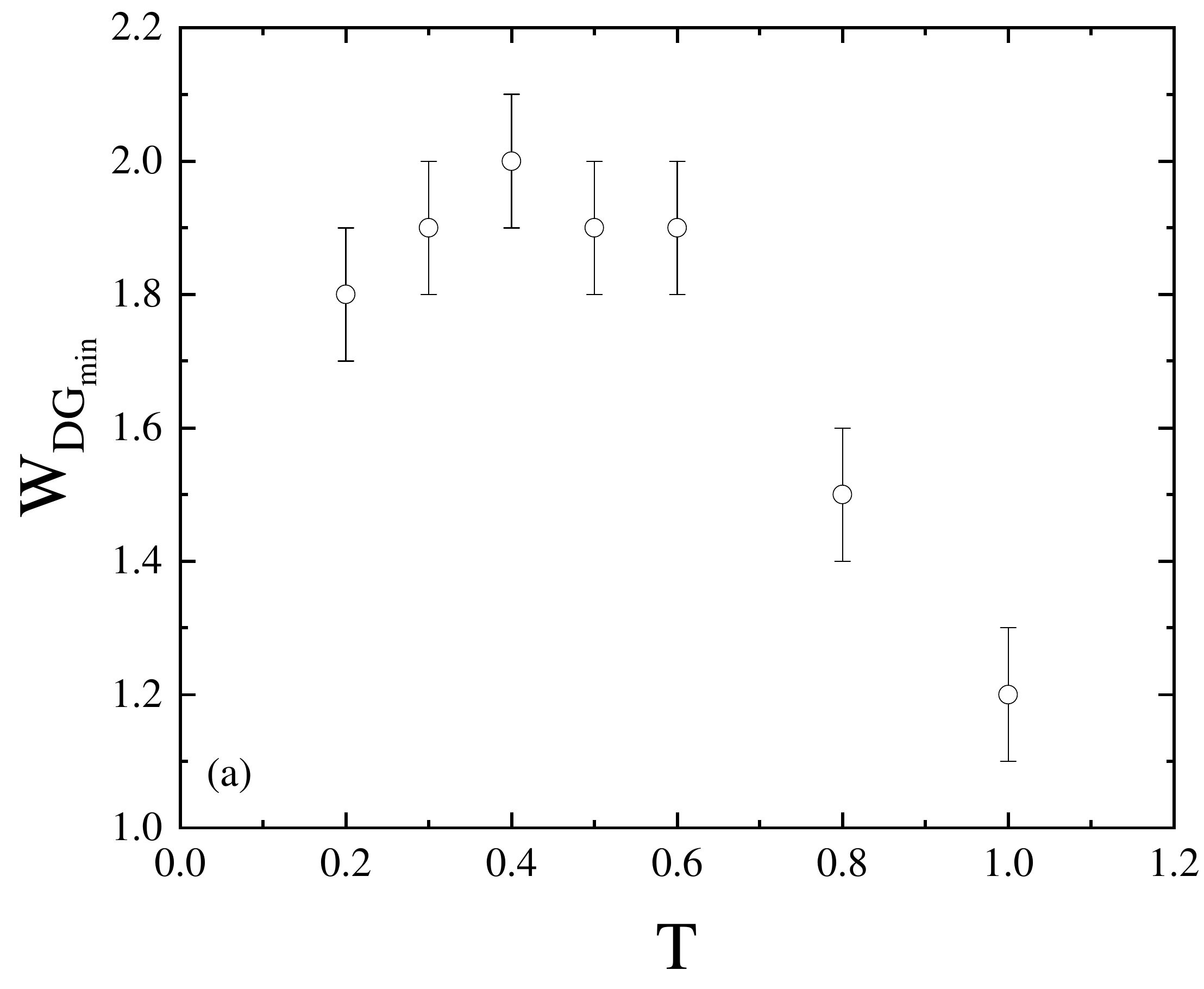}
	\includegraphics[width=80mm]{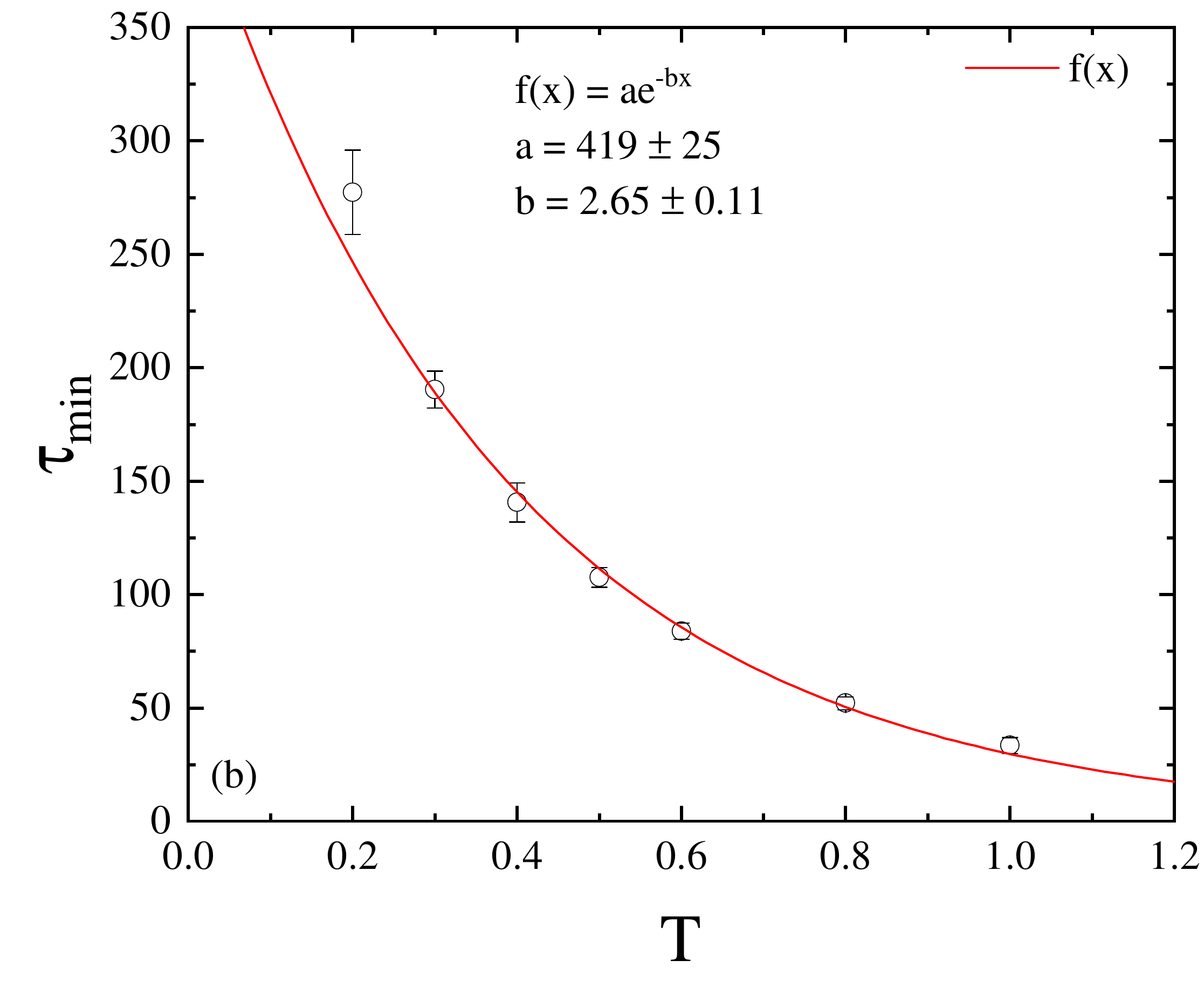}
	\caption{\label{minimums}
	Temperature variation of $W_{\rm DG_{\rm min}}$ (a) and $\tau_{\rm min}$ (b), as obtained from Fig.~\ref{tau_D-Gauss}. $W_{DG_{min}}$ are presented with a kind of systematic error as presented in the case of $T_L^*$. Regarding the fitting shown in (b), we note that $\chi^2/{\rm DOF} = 1.126$  (${\rm DOF} = 5$) so that $Q \sim 34\%$.}
\end{figure}

The behavior of the reversal time was also investigated for the case of a Gaussian of the anisotropy
\begin{equation}
\label{eq:gaussian}
\mathcal{P}_{\rm G}(D_i) = \frac{1}{\sqrt{2\pi\sigma^2}}e^{-\frac{D_i^2}{2\sigma^2}},
\end{equation}
with mean $\mu = 0$ and standard deviation $\sigma = W_{\rm DG}$, see Fig.~\ref{tau_D-Gauss}. Firstly, in panel (a) the evolution of the magnetization for a typical $L = 50$ sample at  $T = 0.2$ is shown for three strengths of the Gaussian distribution, namely $W_{\rm DG} = 0.4, 1.6$, and $2.8$~\cite{box-muller}. A few comments are in order: The reversal time at higher temperatures behaves as if the system is isotropic, similarly to the results of Fig.~\ref{tau_D-bi} for the bimodal distribution of the anisotropy. 
On the other hand, as the temperature is lowered in Fig.~\ref{tau_D-Gauss}(b) we record the appearance of a minimum in the lifetime, $\tau_{\rm min}$, at a certain value of $W_{\rm DG}$, denoted by $W_{\rm DG_{\rm min}}$. Since $W_{\rm DG}$ is varied in steps of $0.1$, the maximum error involved in the determination of $W_{\rm DG_{\rm min}}$ is $\delta W_{\rm DG_{\rm min}}\sim 0.2$. On the other hand the error $\delta\tau_{\rm min}$ are comparable to the size of data points. The variation of both $W_{\rm DG_{\rm min}}$ and $\tau_{min}$ with temperature is depicted in Fig.~\ref{minimums}. 
Although a non-monotonic behavior is observed for $W_{\rm DG_{\rm min}}$ in Fig.~\ref{minimums}(a), the data for $\tau_{\rm min}$ seem to follow nicely an exponential decay as shown by the red solid line in Fig.~\ref{minimums}(b). Some additional results and analysis for the case of a double Gaussian distribution of the anisotropy is given in Appendix~\ref{dG-distribution}.

\section{Conclusions}
\label{sec:conclusions}

\begin{figure}[htbp]
	\includegraphics[width=79mm]{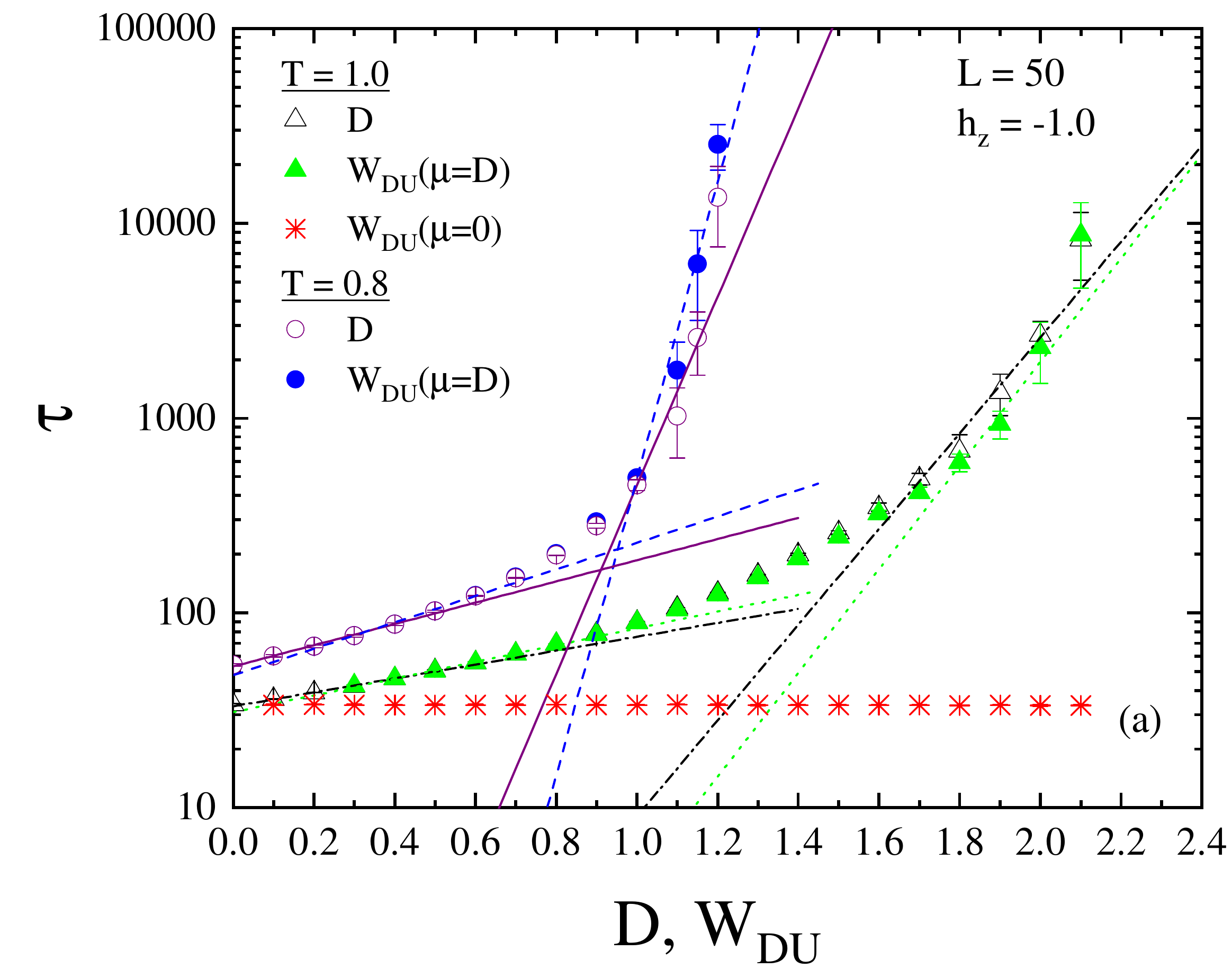}
	\includegraphics[width=76mm]{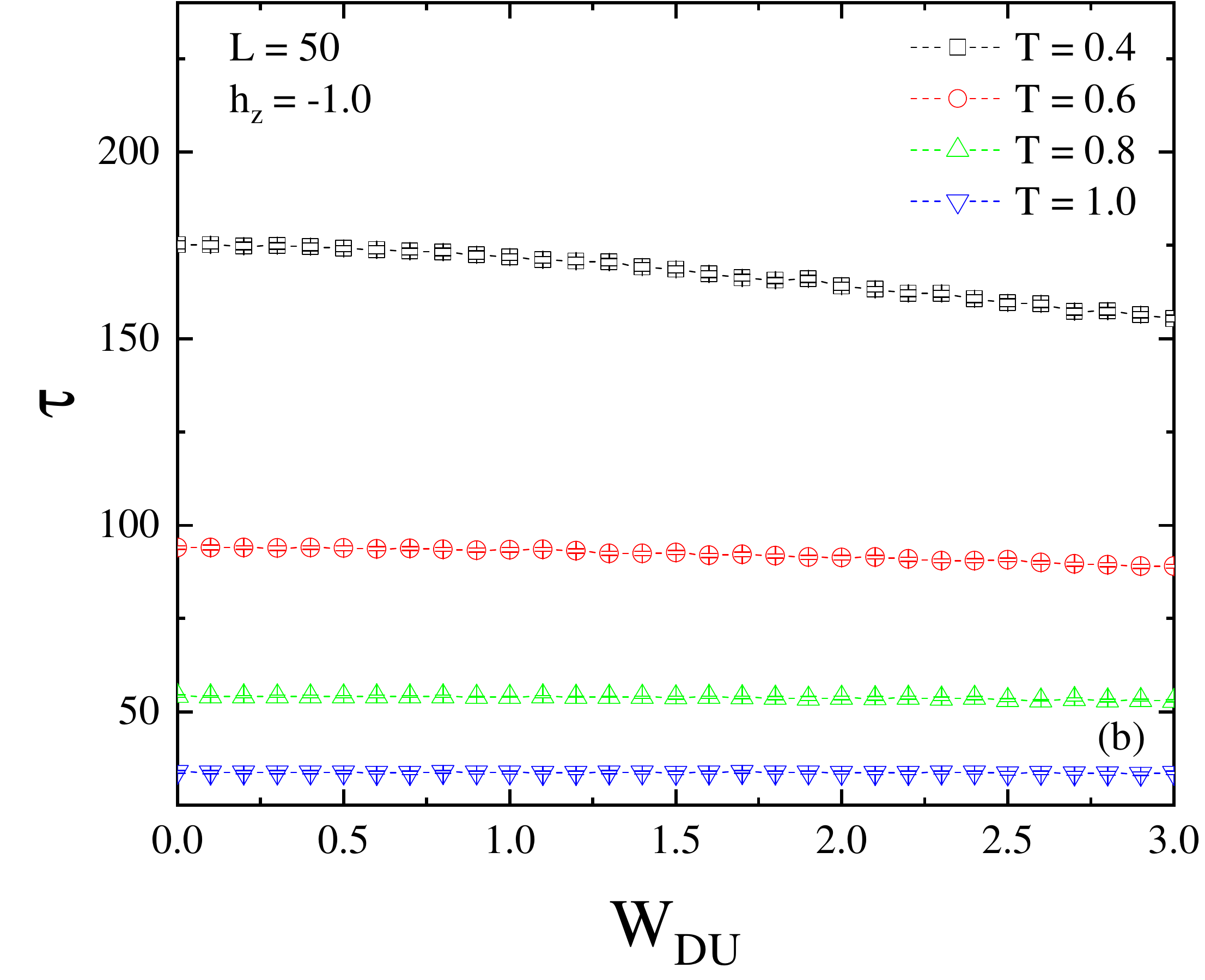}
	\caption{\label{tau-D}
  (a) Variation of reversal time (in logarithmic scale) with: (i) uniform $D$ (purple circles and black triangles), (ii) uniform random distribution $\mathcal{P}(D_i)= 1/W_{\rm DU}$ with mean $\mu=0$ (red stars), and (iii) similar to (ii) but with mean $\mu = D$ (blue circles and green triangles). (b) Reversal time vs. the strength of the uniform random distribution of anisotropy of case (ii) at different temperatures. The different lines are simple exponential fits of the form $f(x) = a \exp(bx)$, as indicated in Tab.~\ref{tab:table1}, for cases (i) and (iii) in the low- and high-$D$ regimes, marking the crossover in the behavior of $\tau$. Indeed, we observe that the position of the crossover point decreases with decreasing temperature. All results correspond to a system with linear size $L = 50$, $h_{z} = -1$, and were averaged over $50$ realizations.}
\end{figure}

The motivation of the current work was to thoroughly investigate the role of disorder (in the form of randomly distributed anisotropy) in metastability phenomena of continuously symmetric ferromagnetic spin systems. In this respect, the three-dimensional classical Heisenberg ferromagnet was used as a platform model to study, by means of Monte Carlo simulations, the behavior of reversal of the magnetization under the presence of a randomly distributed anisotropy. Some additional results were also presented for the combined presence of longitudinal and transverse fields. Although the case of uniform anisotropic Heisenberg model has been widely used in the past, both in theoretical~\cite{bortz72, landau78, dmitriev04, berger08, selke09, albayrak17} as well as experimental~\cite{berger13,berger17} investigations, the scenario of nonuniform (over the space) distribution of anisotropy has not been considered before. The main outcome of our analysis here is that the behavior of the reversal time depends significantly on the statistical distribution of the anisotropy. Another interesting feature emerging from our work is the crossover in the behavior of the reversal time between the low- and high-anisotropy regimes, which is mathematically formalized by the exponential fittings shown in Fig.~\ref{tau-D}(a). These interconnections may be employed in the future in a reversed approach, providing a new route for the identification of the presence and form of the disorder distribution by studying the underlying metastable behavior, \emph{i.e.}, the thermal variation of the switching time. Finally, as it has been shown in the seminal papers by Aharony and collaborators, see Refs.~\cite{aharony78,aharony83} and references therein, at low temperatures, random uniaxial anisotropies
generate local random fields. Of course, for the present case, the situation is
closer to the Ising limit because the anisotropy direction is always
along the $z$-direction. Nevertheless, according to the Imry-Ma criterion~\cite{imry75}, one expects the destruction of long-range order above a critical value of disorder. A detailed study in the absence of field (equilibrium behavior) in order to confirm the destruction of this long-range order is another theoretically motivating endeavor that we plan to investigate in a future project.

\begin{acknowledgments}
M. Naskar acknowledges SVMCM scholarship for financial support. M. Acharyya acknowledges the FRPDF grant of Presidency University. Some of the numerical calculations reported in this paper were performed at T\"{U}B\.{I}TAK ULAKBIM (Turkish agency), High Performance and Grid Computing Center (TRUBA Resources).
\end{acknowledgments}


\newpage 

\begin{table*}
\small
  \caption{Fitting parameters corresponding to Fig.~\ref{tau-D}(a).}
  \label{tab:table1}
  \begin{tabular*}{\textwidth}{@{\extracolsep{\fill}}llllll}
    \hline
    Temperature (T) & Disorder distribution & $D$-regime &   $\chi^2/{\rm DOF}$ & Q $(\%)$ &$b\pm \delta b$ \\
    \hline
    0.8 & D & higher & 1.62 & 20 &11.15 $\pm$ 2.31 \\
        & D & lower & 1.69 & 15 &1.25 $\pm$ 0.04\\
        & $W_{DU}(\mu = D)$ & higher & 2.13 & 12&17.52 $\pm$ 1.94 \\
        & $W_{DU}(\mu = D)$ & lower & 1.49 & 20& 1.56 $\pm$ 0.08 \\
    1.0 & D & higher & 0.95 & 41 & 5.66 $\pm$ 0.51 \\
        & D & lower & 0.96 & 42 & 0.82 $\pm$ 0.02 \\
        & $W_{DU}(\mu = D)$ & higher & 2.13 & 12& 6.13 $\pm$ 1.49 \\
        & $W_{DU}(\mu = D)$ & lower & 1.49 & 20 & 0.99 $\pm$ 0.02\\  
    \hline
  \end{tabular*}
\end{table*}

\newpage

\appendix

\section{Numerical details}
\label{numerics}

The key points of the implemented algorithm are briefly described below: The initial spin configuration ($S_i^x, S_i^y, S_i^z$) consists of all spins parallel to the $z$-direction ($0,0,1$). At a particular temperature, $T$ the system is allowed to follow the Metropolis dynamics in order to change the spin configuration. For any fixed set of ($h_{z}$, $D$) values a lattice site $i$ is chosen randomly. Let us denote the spin vector at this site by ${\bf S_i}$ and the energy of the system $\mathcal{H}$, as given by the Hamiltonian. A test spin vector, say ${\bf S'_i}$, is then chosen (for a trial move) at any direction as follows: two different (uncorrelated) random numbers $r_1$ and $r_2$, uniformly distributed between $-1$ and $+1$, are chosen in such a way that $R^2 = r_1^2 + r_2^2 \leq 1$. Obviously, the set of ($r_1$, $r_2$) values for which $R^2 > 1$ is rejected. Now, if $S_i^x$ and $S_i^y$ are taken as $S_i^x = 2ur_1$ and $S_i^y = 2ur_2$, then $S_i^z$ can be expressed as $S_i^z = 1- 2R^2$, setting $u = \sqrt{1-R^2}$. For this choice of ${\bf S'_i}$ at the site $i$ the new energy is given by $\mathcal{H}' = -J\sum_{\langle ij\rangle} {\bf {S'}_i} \cdot {\bf S_j} - \sum_i D_i ({S'}_i^z)^2 - {\bf h} \sum_i {\bf {S'}_i}$. Thus, the energy change associated with this change in the direction of the spin vector from ${\bf S_i}$ to ${\bf S'_i}$ is given by $\Delta \mathcal{H} = \mathcal{H}'- \mathcal{H}$. At this stage, the Monte Carlo method decides whether the trial move is acceptable or not. 

Its probability follows the Metropolis rate~\cite{binder10}
\begin{equation}
\label{eq:metropolis}
W({\bf S_i} \rightarrow {\bf S'_i}) = {\rm Min}{\Big [}1,{\rm exp}{\big (}-\frac{\Delta \mathcal{H}}{k_B T}{\big )}{\Big ]},
\end{equation}
where $k_{\rm B}$ is the Boltzmann constant and compares the outcome to a random number uniformly distributed between zero and one. If this number does not exceed $W$ then the move ${\bf S_i} \rightarrow {\bf S'_i}$ is accepted and the spin vector ${\bf S_i}$ gets updated. In our numerical experiments, $L^3$ such spin updates define one Monte Carlo step per site, which also sets the time unit of our simulations. 
We also fix $J = k_{\rm B} = 1$ to properly set the temperature scale.

\section{On the quest of finite-size effects} 
\label{finite-size-effects}

\begin{figure}[htbp]
	\includegraphics[width=80mm]{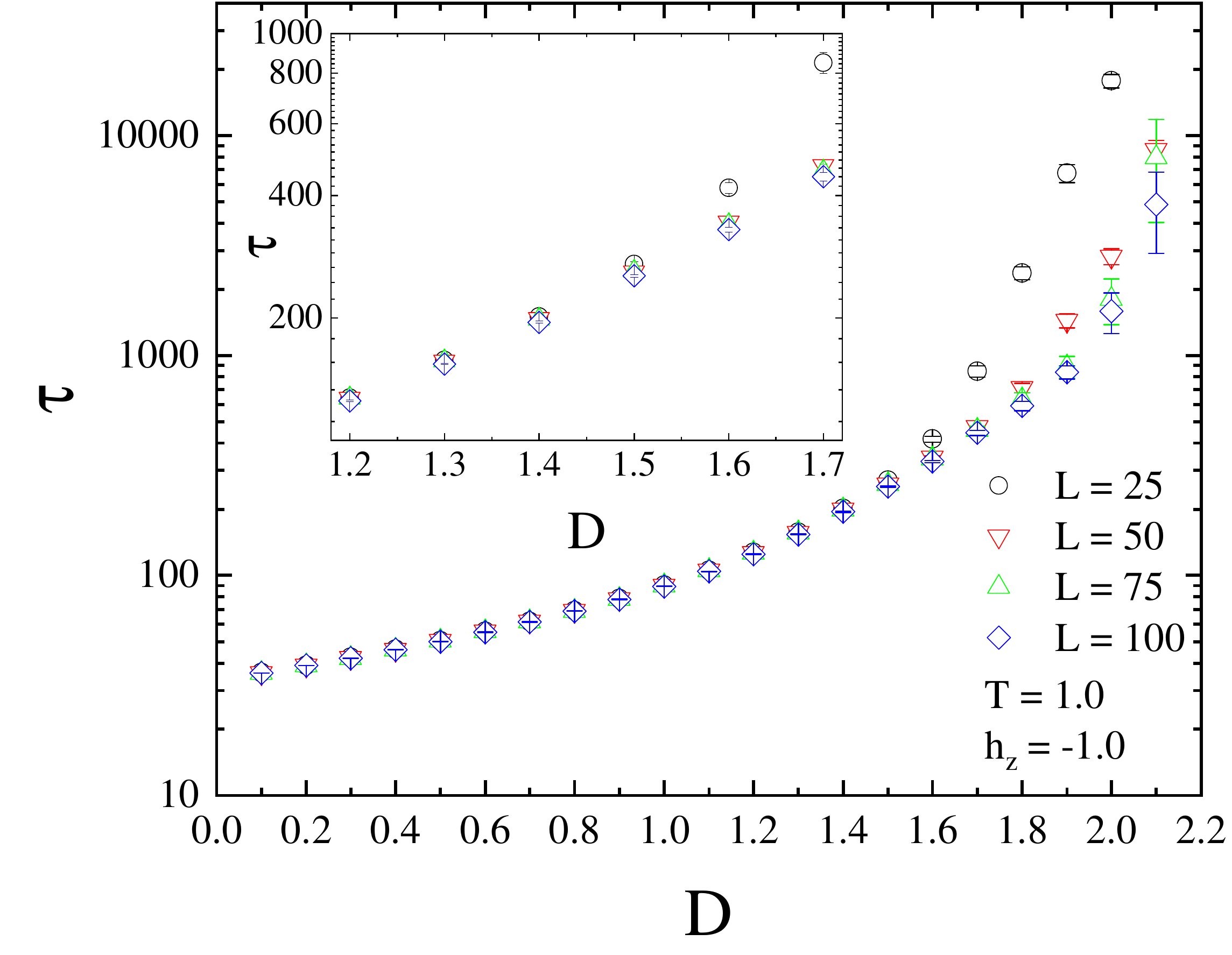}
	\caption{\label{finitesize}
	Variation of reversal time with uniform $D$. Numerical data for $4$ system sizes are shown, namely $L= 25$, $50$, $75$, and $L = 100$, averaged over $1000$, $500$, $50$, and $30$ samples, respectively. All curves are obtained at the selected temperature $T = 1.0$ with $h_z = -1$. The inset is an enlargement of a small section of the curve, at enough magnification so that the errors and, hence, the small deviations for different $L$ can be seen.}
\end{figure}

We present here additional extensive simulations in an attempt to investigate the presence of possible finite-size effects on the variation of reversal time with uniform $D$ (part of Fig.~\ref{tau-D}(a)). Our findings shown in Fig.~\ref{finitesize} below indicate that the reversal time decreases with a converging behavior with increasing system size, but with no scaling properties being detected. In particular for systems with $L \geq 75$ the numerical data become almost $L$-independent, with some minor deviation being present.

\section{The case of the double Gaussian distribution}
\label{dG-distribution}

\begin{figure}[htbp]
	\includegraphics[width=79mm]{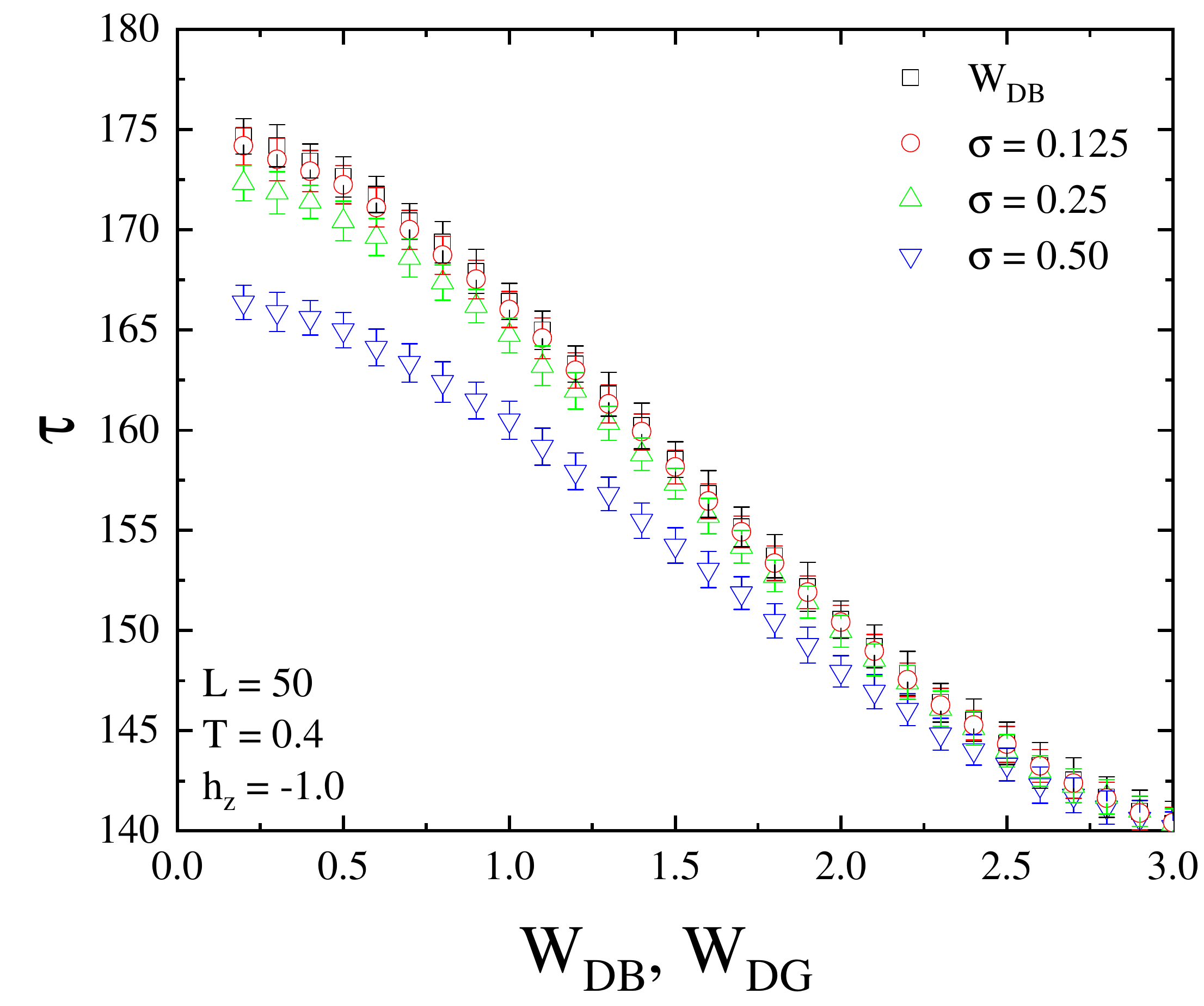}
	\caption{\label{comparison}
		Reversal time vs. the strength of (i) the bimodal distribution of anisotropy~(\ref{eq:bimodal}) for $p = 0.5$, and (ii) the double Gaussian distribution of the anisotropy~(\ref{eq:dgaussian}) with $\sigma=0.5$, $0.25$, and $\sigma = 0.125$ at $T = 0.4$. All results correspond to a system with linear size $L = 50$, $h_{z} = -1$, and were averaged over $150$ realizations.}
\end{figure}

It is also interesting to consider a double Gaussian distribution of the anisotropy
\begin{equation}
\label{eq:dgaussian}
\mathcal{P}_{\rm dG}(D_i) = \frac{1}{\sqrt{8\pi\sigma^2}}\big[e^{-\frac{(D_{i}-\mu)^{2}}{2\sigma^{2}}}+
  e^{-\frac{(D_{i}+\mu)^{2}}{2\sigma^{2}}}\big],
\end{equation}
where now the mean of the distribution is $\mu= W_{\rm DG} / 2$. In the limit $\sigma \rightarrow 0$ we expect the behavior of the reversal time to be equivalent to that under the presence of the discrete bimodal distribution~(\ref{eq:bimodal}), studied above. This is indeed well verified from the numerical data of Fig.~\ref{comparison}.

\newpage


\end{document}